\documentclass[11pt]{article}
\usepackage{latexsym,amsmath,amssymb, graphicx,subfig,fullpage}
\usepackage{tikz}
\usetikzlibrary{shapes,arrows}

\tikzstyle{blockc}  =  [draw, fill=blue!20, rectangle,  minimum  height=2em,  minimum  width=4em]
\tikzstyle{blocko}  =  [draw, fill=green!20, rectangle,  minimum  height=2em,  minimum  width=4em]
\tikzstyle{blockp}  =  [draw, fill=red!20, rectangle,  minimum  height=2em,  minimum  width=4em]
\tikzstyle{sum} = [draw, fill=gray!20, circle, node distance=1cm]
\tikzstyle{input} = [coordinate] \tikzstyle{output} = [coordinate]
\tikzstyle{pinstyle} = [pin edge={to-,thin,black}]

\DeclareMathAlphabet{\mathcal}{OMS}{cmsy}{m}{n}

\newcommand{\ltwo}{\mathbb{Z}}
\newcommand{\opA}{\mathcal{A}}
\newcommand{\opB}{\mathcal{B}}
\newcommand{\opP}{\mathcal{P}}
\newcommand{\mathQ}{\mathcal{Q}}

\newcommand{\dom}{\mathcal{D}}
\newcommand{\mathK}{\mathcal{K}}
\newcommand{\opC}{\mathcal{C}}
\newcommand{\linop}{\mathcal{L}}
\newcommand{\ctrlu}{u}
\newcommand{\linf}{\mathcal{L}_{\infty}}
\newcommand{\mathW}{\mathbb{W}}
\newcommand{\lon}{\mathcal{L}_1}
\newcommand{\mathR}{\mathbb{R}}
\newcommand{\dwdt}{\dot{w}}
\newcommand{\citep}{\cite}
\newcommand{\staropT}{\mT\star}
\newcommand{\staropTnorm}{\|\mT\ast\|}
\newcommand{\mT}{\mathcal{T}}
\newcommand{\projfn}{\pi}
\newcommand{\bigw}{\boldsymbol{w}}

\newtheorem{assm}{Assumption}
\newtheorem{prop}{Proposition}
\newtheorem{thm}{Theorem}
\newtheorem{lem}{Lemma}

\newtheorem{defn}{Definition}
\newtheorem{remark}{Remark}

\title{\bf \Large Transient and Asymptotic Properties of Robust Adaptive Controllers in the Presence of Non-Coercive Lyapunov Functions}
\author{Aditya A. Paranjape, Vivek Natarajan and Supratim Ghosh %
\thanks{AP and SG are with TCS Research, a part of Tata Consultancy
Services Limited, India. Email: \{aditya.paranjape@tcs.com; supratim.ghosh2@tcs.com\}. VN is with the Systems and Control Engineering Group at the Indian Institute of Technology Bombay 
in Mumbai, India. Email: vivek.natarajan@iitb.ac.in.}}

\begin{document}
\maketitle
\thispagestyle{empty}
\pagestyle{empty}

\begin{abstract}
Adaptive control architectures often make use of Lyapunov functions to design
adaptive laws. We are specifically interested in adaptive control methods, such
as the well-known $\lon$ adaptive architecture, which employ a parameter 
observer for this purpose. In such architectures, the observation error plays a critical role 
in determining analytical bounds on the tracking error as well as robustness. In this
paper, we show how the non-existence of coercive Lyapunov operators can impact the analytical bounds, and with it
the performance and the robustness of such adaptive systems.
\end{abstract}

\section{Introduction}
Lyapunov equations with non-coercive solutions are a peculiar feature of infinite dimensional systems \cite{dat68,paz72}. 
In a finite dimensional setting, the Lyapunov equation corresponding to a Hurwitz matrix yields a positive definite solution. 
In semilinear systems of the form $\dot{w} = Aw + f(w)$, where $A$ is Hurwitz, one can use this solution to determine permissible 
bounds on $f(w)$ as well as the associated bounds on the solution $w(t)$.

In infinite dimensional systems, the impact of non-coercivity
can be felt on the nature of bounds that can be derived for 
$w(t)$; see \cite{mir19non,jac20non} for 
example. There are ways to get around the non-coercivity,
by invoking additional assumptions on the system (e.g., 
a stronger form for the Lyapunov equation \cite{cur03}) or 
delicate fictitious modifications which aid the
derivation of a coercive Lyapunov function \cite{wen89}. 

In this paper, we will consider robust adaptive control of 
systems of semilinear partial differential equations (PDEs) of the form $\dot{w}(t) = \opA w(t) + \opB u(t)+ f(w),~y(t) = \opC w(t)$,
where $w(t)$ denotes the system state, $u(t)$ is the control input, and $y(t)$ is the output. The operators 
$\opA$, $\opB$ and $\opC$ are the state, control, and output operators, respectively.
Coercive Lyapunov functions feature prominently in the derivation of adaptive laws, and help 
ensure appropriate bounds on the tracking error \cite{nat12,par16cdc,par18dac}. Our objective is to determine how the
guaranteed bounds change in the absence of a coercive solution to the usual, unmodified Lyapunov equation.

\begin{figure}[htb]
\centering
\begin{tikzpicture}[auto, node distance=2cm,>=latex']
    \node [input, name=input] {};
    \node [sum, right of=input] (sum) {};
    \node [blockc, right of=sum] (filter) {$H(s)$};
    \node [output, right of=filter,node distance=3cm] (virtualnode) {};
    \node [output, right of=virtualnode, node distance=2cm] (virtualnode2) {};
    \node [blocko, below of=filter,node distance=1.5cm] (phalf) {Particular Half};
    \node [blockp, below of=virtualnode, node distance=1.5cm] (sys) {Plant};
    \node [blocko, below of=sys, node distance=1.5cm] (hhalf) {Homogeneous Half};
    
    \draw [draw,->] (input) -- node {$r(t)$} (sum);
    \draw [->] (sum) -- node {} (filter);
    \draw [-] (filter) -- node {$u(t)$} (virtualnode);
    \draw [-] (virtualnode) -- node {} (virtualnode2);
    \draw [->] (virtualnode) -- node {} (sys);
    \draw [->] (sys) -- node {$v(t)$} (phalf);
    \draw [->] (phalf) -| node [pos=0.98] {$-$}
    	node [near end] {$\hat{y}_p(t)$} (sum);
    \draw [->] (virtualnode2) |- node{} (hhalf);
    \draw [->] (hhalf) -| node {$\hat{v}_h(t)$} (phalf);
\end{tikzpicture}
\caption{A block diagram of the DAC framework, with the subscripts {\it p} and {\it h} denoting
signals from the particular and homogeneous components. The symbols $v(t)$, $y(t)$, and $r(t)$ denote the system state, output and reference signal, respectively.}
\label{fig:lure}
\end{figure}
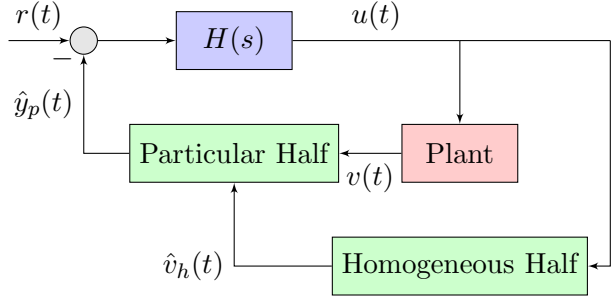

\subsection{Contribution}
In this paper, we examine the effects of non-coercive Lyapunov functions on the performance and stability of
semilinear infinite dimensional systems controlled by an adaptive controller based on the $\lon$ philosophy
\cite{hovbook}. In particular, we consider a semilinear system with unmatched uncertainties, and 
a dyadic adaptive architecture based on \cite{par18dac}, see Fig.~\ref{fig:lure}.

It has been shown previously \cite{nat12,par18dac} how a coercive Lyapunov function helps derive
tight bounds on the tracking performance and the control inputs. In this paper, we extend the 
analysis to derive weaker bounds when a coercive Lyapunov function cannot be found.

Although we consider a specific dyadic adaptive architecture in this paper, our conclusions or results
can be extended to other adaptive architectures such as model reference adaptive control (MRAC) where 
Lyapunov analysis is used to derive the adaptive laws and prove that the error between the reference
model and the system is suitably bounded.

The paper is organized as follows. We introduce the mathematical preliminaries in Sec.~\ref{sec:prelim},
and the problem formulation in Sec~\ref{sec:pf}. In 
Sec.~\ref{sec:udesign}, we present the design of 
the control law. In Sec.~\ref{sec:obserr}, we 
show the boundedness of the observation error.
We discuss closed-loop stability and model-following in Sec.~\ref{sec:stab}.

\section{Preliminaries}\label{sec:prelim}
\subsection{Spaces, operators and norms}
\begin{defn}[$\linf$ and $\lon$ norms]
Given $q(t) \in \mathR^n$ with components $q_i(t)$ ($1 \leq i \leq n$), we define 
\begin{eqnarray*} 
\|q(t)\|_\infty & = &\max_{1 \leq i \leq n} |q_i(t)|,~\|q\|_{\linf} = {\rm ess} \sup_{t \geq 0} \|q(t)\|_{\infty} \\
\|q\|_{\linf,\tau} &=& {\rm ess} \sup_{0 \leq t \leq \tau} \|q(t)\|_\infty
\end{eqnarray*}
If $\|q\|_{\linf} < \infty$, then we denote $q \in \linf^n$. The $1$-norm of a matrix $F:\mathR^m \to \mathR^n$ is defined as 
$\|F\|_{1} = \sup_{\|p\|_{\infty} = 1} \|Fp\|_{\infty},~p\in\mathR^m$. The $\lon$ norm of a linear operator $\mathcal{F}:\linf^m \mapsto \linf^n$ is defined as $\|\mathcal{F}\|_{\lon} = \sup_{\|q\|_{\linf} = 1} \|\mathcal{F}q\|_{\linf},~q\in\linf^m$
\end{defn}

The spatial domain of interest in this paper is the closed interval $[0,\,L]$ for some finite $L > 0$. 
Let $\ltwo = \mathcal{L}_2([0,\,L],\,\mathbb{R}^n)$ denote the Hilbert space of square integrable functions with the usual inner product and norm. 

\begin{defn}\label{def:norm}
We define $\mathW$ as the space of $\ltwo$-valued functions on $\mathbb{R}_{\geq 0}$ satisfying
${\rm ess}\,\sup_{t\geq 0} \|w(t)\|_{\ltwo} < \infty$. 
The space $\mathW$ is a Banach space with the norm
$\|w\|_{\mathW} = {\rm ess}\sup_{t \geq 0}\|w(t)\|_{\ltwo}$. For $\tau \geq 0$, we define 
the truncated norm given by
$\|w\|_{\mathW,\tau} = {\rm ess}\sup_{0\leq t \leq \tau} \|w(t)\|_{\ltwo}$ and the associated Banach space by
$\mathW_\tau$.
\end{defn}

\begin{defn}\label{defn:linop}
We denote a linear operator between spaces $X$ and $Y$
by $\linop(X,Y)$.
\end{defn} 

\begin{defn}[\citep{pazbook}, Definition 1.1, Ch. 6]\label{lem:voc}
Consider a system $\dwdt = \opA w + f(t,w)$, $w(t=0) = w_0 \in \ltwo$, where $\opA$ is the 
infinitesimal generator of a $C_0$ semigroup $\mT(t)$ and $f(t,w)$ is continuous in $t$ and satisfies a Lipschitz condition in $w$. The mild solution $w(t)$ is given by
\begin{equation}
w(t) = \mT(t)w_0 + \int_0^t \mT(t-\tau)f(\tau,w(\tau))\,d\tau,
\label{eq:ivp0mild}
\end{equation}
\end{defn}
\begin{defn}[Convolution]\label{def:star}
Given a $C_0$ semigroup $\mT(t)$ and $\tau \geq 0$, 
we define the operator $\staropT(t)(\tau): 
\mathW_\tau \mapsto \mathW_\tau$ as
$\staropT (\tau) f = \int_0^\tau \mT(\tau-s) f(s,w(s))\,ds~\forall\,f \in \mathW_\tau$.
We define  the induced norm
$\staropTnorm \triangleq {\rm ess}\,\sup_{(t \geq 0)} \|\staropT(t)\|$.
\end{defn}

We recall the following result from \citep{pazbook} for solutions of initial 
value problems in Definition~\ref{lem:voc}.
\begin{thm}[Theorems 6.1.4, 6.1.5, \citep{pazbook}] \label{thm:paz} Let $\opA$ 
be the infinitesimal generator of a $C_0$ semigroup $\mT(t)$ on the Hilbert space $\ltwo$. 
If $f:[0,\,T]\times \ltwo \to \ltwo$ is continuously differentiable with 
respect to both arguments, for 
$T > 0$, then the mild solution \eqref{eq:ivp0mild} is a classical solution of the initial value problem in Definition~\ref{lem:voc}
for $t \in [0,\,T]$. If the solution exists only up to $T_{\max} < T$, then 
$\|w(t)\|_{\ltwo} \to \infty$ as $t \to T_{\max}$.
\end{thm}

Next, we define the projection operator \cite{lavretsky2011projection} which
will be used for constructing the adaptive laws. 
Let $\projfn:\,\mathbb{R}^k \to \mathbb{R}$ be defined by
$$
\projfn(\alpha) \equiv \projfn(\alpha; \kappa,\epsilon) = \frac{\langle \alpha,\,\alpha\rangle - \kappa^2}{\epsilon \kappa^2},~~\alpha \in \mathbb{R}^k,~\kappa \in \mathbb{R}^{+}
$$
The number $\epsilon \in \mathbb{R}^{+}$ 
is chosen to be arbitrarily small. The Fr\'echet derivative of $\projfn$ at 
$\alpha_1 \in \mathbb{R}^k$ is denoted by $\projfn'(\alpha_1) \in \mathbb{R}^k$ and it satisfies
$$
\langle \projfn'(\alpha_1),\,\alpha_2\rangle = \frac{2\langle \alpha_1,\,\alpha_2 \rangle}{\epsilon K^2}~~\forall \alpha_2 \in \mathbb{R}^k
$$

\begin{defn}\label{defn:proj}
The projection operator ${\rm Proj}:~\mathbb{R}^k \times \mathbb{R}^k \to \mathbb{R}^k$ is defined as
\begin{equation}
{\rm Proj}(\alpha_1,\alpha_2) \! = \!\begin{cases} \alpha_2,~{\rm if}~\projfn(\alpha_1) \leq 0 ~{\rm or}~
\langle \projfn'(\alpha_1),\,\alpha_2 \rangle \!\leq\! 0 \\
\alpha_2 \!-\! \frac{\projfn'(\alpha_1)}{\|\projfn'(\alpha_1)\|_2} \!\times\! 
\left\langle \frac{\projfn'(\alpha_1)}{\|\projfn'(\alpha_1)\|_2} ,\,\alpha_2 \right\rangle \projfn(\alpha_1), \\ {\rm otherwise}
\end{cases}
\label{eq:projdef}
\end{equation}
\end{defn}

\begin{lem}[Lemma 9 in \cite{lavretsky2011projection}]\label{lem:lavr}
 Let $\Omega_0$ and $\Omega_1$ denote the convex
sets satisfying 
$$
\Omega_0 = \left\{\alpha~|~\pi(\alpha) \leq 0 \right\},~~
\Omega_1 = \left\{\alpha~|~\pi(\alpha) \leq 1 \right\}
$$
Suppose that $\alpha_1^\ast \in \Omega_0$. Then, for all $\alpha_1,\,\alpha_2 \in 
\mathbb{R}^k$, 
$(\alpha_1 - \alpha_1^\ast)\left({\rm Proj}(\alpha_1,\,\alpha_2) - \alpha_2\right) \leq 0$.
Moreover, the solution of the initial value problem 
$\dot{\alpha}_1 = {\rm Proj}\left(\alpha_1,\,\alpha_2\right),~\alpha_1(0) = \alpha_{10}$, has
the property that if $\alpha_{10} \in \Omega_1$, then $\alpha_1(t) \in \Omega_1$
for all $t$. 
\end{lem}

\subsection{Stability}
We will need the following weaker notion of asymptotic stability, in addition
to the more usual notions of stability.
\begin{defn}\label{defn:almostasymp}
We say that a function $f:\mathbb{R_{\geq0}} \to \mathbb{R}$ converges to $0$ {\em almost asymptotically} if 
$$
\lim_{n\to\infty} f(nx) = 0~~\textrm{for almost all}\,x\geq0
$$
\end{defn}

\begin{lem}[\cite{les10}, Theorem 1]\label{lem:almostasymp}
If $f(t) \in \mathcal{L}_2(\mathbb{R_{\geq0}} ,\mathbb{R})$, then $f(t)$ converges to $0$ almost asymptotically, 
in the sense of Definition~\ref{defn:almostasymp}.
\end{lem}

Consider the abstract system $\dot{z} = \opA_m z + g(z),~z(0) = z_0$, where $z_0 \in \dom(\opA_m)$ and $\opA_m$ is the infinitesimal generator of 
an exponentially stable semi-group $\mT(t)$ and 
$g$ is satisfies a Lipschitz condition in $z$. 
The following lemma asserts the existence of a Lyapunov function corresponding to $\opA_m$.
\begin{assm}[based on \cite{nat12}]\label{assm:lyap}
Let $\mathQ > 0$ be a self-adjoint, boundedly invertible operator on 
$\ltwo$; i.e., $\mathQ^{-1} \in \linop(\ltwo)$ and with $\dom(\opA_m)
\subseteq \dom(\mathQ)$. We assume that there exists 
$\opP \in \linop(\ltwo)$ with $\opP > 0$ such that,
\begin{eqnarray}
\nonumber \hspace{-7mm} & & \langle \opA _m z,\opP z \rangle_{\ltwo} 
\!+\! \langle \opP z,\opA _m z \rangle_{\ltwo} \leq -\!\langle z,\mathQ \,z \rangle_{\ltwo}, \\
\hspace{-7mm} & & \forall~z \in \dom(\opA _m)
\label{eq:lyap}
\end{eqnarray}
\end{assm}
We note that a solution $\opP \in \linop(\ltwo)$ exists if 
$\mathQ \in \linop(\ltwo)$ \cite{cur95}. 
\begin{remark}\label{rem:lyap01}
If $\opA_m + \opA_m^\ast < 0$ or if $\opA_m$ is the infinitesimal generator of $C_0$ group \cite{paz72}, then it is possible
to find $\mathQ$ satisfying Assumption~\ref{assm:lyap} such
that the $\opP \in \linop(\ltwo)$ is coercive.
\end{remark}

\section{Problem Formulation}\label{sec:pf}
This paper is concerned with the control of systems of semilinear infinite 
dimensional systems described by
\begin{eqnarray}
\dot{w}(t) = \opA w(t) + \opB u(t) + \alpha f(w),~ y(t) = \opC w(t)
\label{eq:iom}
\end{eqnarray}
where $w(t) \in \ltwo$ and $u(t) \in \mathbb{R}$, $\opB \in \linop(\mathbb{R},\mathbb{R})$ 
and $\opC \in \linop(\ltwo,\mathbb{R})$.
The control objective is to design $u(t)$ so that the output $y(t)$ tracks a reference signal $r(t)$, and the resulting closed-loop system
is stable and robust in the sense of $\linf$.

\begin{assm}\label{assm:lipschitz}
The nonlinearity $f(w)$ is a known $C^1$ function of $w$, while $\alpha \in \mathR^n$ is unknown but
satisfies $|\alpha_{i}| < \nu_\alpha$
for all $i \in \{1,2,\dots,\,n\}$. 
\end{assm}
The analysis in the paper does not require that 
$\alpha_i$ be a constant, and neither is it necessarily restricted to a single ``basis'' function $f(w)$ (see \cite{par12cdc}). This assumption does, however, 
simplify the presentation.

\begin{assm}
The permissible initial conditions are restricted by $\|w_0\|_{\ltwo } < \rho_0$, and
$w _0 \in \dom(\opA)$.
\end{assm}
\begin{assm}[Stabilizability]\label{assm:stabilizability}
There exists $\mathK \in \linop(\ltwo,\mathbb{R})$ such that
$\opA - \opB\mathK$ is the infinitesimal generator of an 
exponentially stable semi-group.
\end{assm}

\begin{lem}\label{lem:lipschitz}
For every $\rho > 0$, there exist constants $\nu_1(\rho)$ and $\nu_2(\rho)$
such that if $\|w \|_{\mathW ,\tau} < \rho$ for some $\tau > 0$, then
$\|f(w )\|_{\mathW,\tau} \leq \nu_1(\rho)\|w \|_{\mathW ,\tau} + \nu_2(\rho)$.
\end{lem}

\section{Control Design}\label{sec:udesign}
\subsection{Control Signal}
Consider the system
\begin{equation}
\dot{w} _h(t) = \opA  w _h(t) + \opB  \ctrlu (t),~y_h = C  w _h \label{eq:ha0}
\end{equation}
which is found by neglecting the nonlinearity in \eqref{eq:iom}. Using 
Assumption~\ref{assm:stabilizability}, we deduce that there exists a bounded stabilizing
gain $\mathK: \ltwo \to U$ such that $\opA - \opB\mathK$ generates an exponentially
stable $C_0$ semigroup. We formalize this as follows.

\begin{defn}
The operator $\opA _m = \opA  - \opB \mathK$ generates an exponentially
stable semigroup $\mT(t)$; i.e., there exist constants $M,\,\beta > 0$ such that $\|\mT(t)\|_i \leq Me^{-\beta t}$. Moreover,
$\staropTnorm$ is bounded.
\end{defn}

Based on our prior work \cite{par14cdc, par16cdc}, we use the following control 
law for the system \eqref{eq:iom}:
\begin{eqnarray} 
\ctrlu(t) &=& -\mathK w (t) - H_Cp(t)
\label{eq:upt} \\
\dot{p}(t) &=&H_A p(t) + H_B \sigma(t),~p(0) = p_0 
\label{eq:pbvp}
\end{eqnarray}
with $H_A$ Hurwitz. The term $\sigma(t)$, on which $p(t)$ depends, will be defined presently. 
The terms $H_C$ and $H_B$ are chosen to satisfy the DC gain condition $\opC (-\opA_m)^{-1}\opB H_C (-H_A)^{-1}H_B = -1$.

The system \eqref{eq:iom} can now be written as
\begin{equation}
\dot{w} (t) = \opA _m w (t) - \opB  H_C p(t) + \alpha f (w (t))
\label{eq:iomc}
\end{equation}
Using the linear term as a pivot, we decompose the system in 
\eqref{eq:iom} into two sub-systems
\begin{eqnarray}
\dot{w} _p &=& \opA_m  w _p + \alpha f (w ),~y_p = C  w _p \label{eq:pa} \\
\dot{w} _h &=& \opA_m  w _h - \opB  H_C p(t),~y_h = C  w _h \label{eq:ha}
\end{eqnarray}
The two systems \eqref{eq:pa} and \eqref{eq:ha} are referred to as the 
{\em particular} and {\em homogeneous} halves, respectively. In the next
section, we will derive an observer for estimating the states; for now, we
use \eqref{eq:pa} and \eqref{eq:ha} to investigate tracking.

If we could choose $\sigma(t) = r(t) - y_p(t)$, we would get that the tracking error $y(t) - r(t) = y_h(t) - \sigma(t)$; therefore, $\sigma(t)$ can serve as the reference
signal for $y_h(t)$. Since $y_p(t)$ is not known, we will choose
\begin{equation}
\sigma(t) = r(t) - \hat{y}_p(t)
\label{eq:sigma}
\end{equation}
where $\hat{y}_p(t)$ is the output of an observer which 
will be designed presently (see \eqref{eq:ppe}).

\subsection{Observer Design}\label{sec:obsdyn}
We use the symbol ``$\wedge$'' to denote observer states, and the subscripts {\it p} and {\it h} to denote states of the particular and the homogeneous halves, respectively. The dynamics of the two halves are given by
\begin{eqnarray}
\hspace{-10mm}&& \dot{\hat{w}} _p \!=\! \opA _m \hat{w} _p + \hat{\alpha}(t)f(w),
~\hat{y}_p = \opC  \hat{w} _p \label{eq:ppe}\\
\hspace{-10mm}&& \dot{\hat{w}} _h \!=\! \opA _m \hat{w} _h - \opB  H_C p(t),~\hat{y}_h = \opC  \hat{w} _h \label{eq:hpe}
\end{eqnarray}
with the initial conditions $\hat{w} _h(0) = w (0)$ and $\hat{w} _p(0) = 0$. 

The predicted values $\hat{\alpha}(t)$ are found using the projection operator (see
\cite{lavretsky2011projection}, \cite{nat12} for details).
\begin{eqnarray}
\nonumber & & \dot{\hat\alpha}_{i}(t) = \gamma\,{\rm Proj}\left(\hat{\alpha}_{i},\,-\langle \opP\tilde{w} (t), f(w)e_i \rangle_{\ltwo} \right),\\
&& |\hat\alpha_{i}(t)| < \nu_\alpha(1 + \epsilon)
\label{eq:pproj}
\end{eqnarray}
where $\epsilon \in \mathbb{R}^{+}$ is arbitrarily small; $\tilde{w}  = \hat{w} _p + \hat{w} _h - w $; 
$\hat{\alpha}_{i}$ ($1 \leq i \leq n$) is the $i^{\rm th}$ component of 
$\hat{\alpha}$, $e_i$ denotes the $i^{\rm th}$ column of the $n \times n$ identity matrix, and $\gamma > 0$ is the 
adaptation gain.

The operator $\opP > 0$ in \eqref{eq:pproj} is found by solving
the Lyapunov equation \eqref{eq:lyap} with $\mathQ $ chosen
as follows:
\begin{eqnarray}
\mathQ  = \begin{cases} -(\opA_m + \opA_m^\ast) & {\rm if }~\opA_m + \opA_m^\ast < 0
 \\
\mathcal{I} & {\rm otherwise}
\end{cases}
\label{eq:Qselect}
\end{eqnarray}
where $\mathcal{I}$ is the identify operator on $\ltwo$. In the first
case, it can be seen that $\opP = \mathcal{I}$, which is coercive.

In summary, the closed-loop system consists of the original system
\eqref{eq:iom}, together with the controller \eqref{eq:upt}, and the 
dyadic observer \eqref{eq:ppe}, \eqref{eq:hpe} and \eqref{eq:pproj}.

\subsection{Well-Posedness}
To analyze the well-posedness of the closed-loop system, we construct the augmented vector $\bigw  = [w ,\,\hat{w} _p,\,
\hat{w} _h,\,p(t)]^{\top} \in \mathbb{V} = \ltwo  \times \ltwo  \times \ltwo  \times \mathR^{n_p}$. The dynamics of $\bigw $ is given by
\begin{eqnarray}
&& \dot{\bigw} (t) = \bar{\mathcal{A}}  \bigw  + \bar{f}(\hat{\alpha}(t),w(t),r(t)) 
\label{eq:big}\\
\nonumber  && w (0) = \hat{w} _h = w _0,~\hat{w} _p(0) = 0,~~p(0) = p_0 \\
\nonumber && \bar{\mathcal{A}} = \left[\begin{matrix} \opA _m & 0 & 0 & -\opB  k_r \\ 0 & \opA _m & 0 & 0 \\
0 & 0 & \opA _m & -\opB  k_r \\  0 & - H_B & 0 & H_A \end{matrix}\right]
\end{eqnarray}
where the exogenous signal $\hat{\alpha}(t)$ is known to be $C^1$ in time. Therefore, it can be 
checked readily that $\bar{f}(\cdot)$ is a $C^1$ function of its arguments. Furthermore, the operator 
$\bar{\mathcal{A}} $ is the infinitesimal generator of a semigroup. We state the following
result without proof, but as a direct application of Thm.~\ref{thm:paz}.
\begin{lem}\label{lem:classical}
There exists $T_{\max} > 0$ such that the system 
\eqref{eq:big} has a unique classical solution $\bigw (t)$ for
$t \in [0,\,T_{\max}]$. Moreover, if $T_{\max} < \infty$, then
$\lim_{t\to T_{\max}}\|\bigw (t)\|_{\mathbb{V}} \to \infty$. 
\end{lem}

\subsection{A Necessary Condition for Tracking} 
\begin{lem}[Necessary condition for tracking]\label{lem:tracknc}
Suppose that $w$ is suitably bounded and we 
design $u(t)$ to ensure that $\hat{y}_h(t)$ tracks $\sigma(t) = r(t) - \hat{y}_p(t)$, where the signals have been
defined in \eqref{eq:sigma}, \eqref{eq:ppe} and \eqref{eq:hpe}.
Then, $y(t)$ tracks $r(t)$ only if $y(t)$ tracks $\hat{y}(t)$.
\end{lem}
The necessary condition stated here is quite obvious, but its role
will become clear in the subsequent analysis. Informally speaking,
when coercivity is lost, it may no longer be possible to prove
asymptotic bounds on the observer states themselves, but one
can prove asymptotic bounds on the observer output.
In the next section, we will prove output error regulation. 

\section{Observer Error Regulation}\label{sec:obserr}
In this section, we derive bounds on the observation error
between $\hat{w} _h$ and $\hat{w} _p$ on the one hand and
$w_p $ and $w _h$, respectively, on the other. Let 
$\hat{w}  = \hat{w} _h + \hat{w}_p$, and let $\tilde{(\cdot)} = \hat{(\cdot)} - (\cdot)$ denote the error between predicted and the actual terms. We have two objectives: derive tight bounds 
on $\tilde{w}$ and $\tilde{y}$, and show that $\tilde{y}$ converges to zero asymptotically if an arbitrarily tight bound 
(in a sense that will become clear presently) cannot be derived.

From \eqref{eq:iomc}, \eqref{eq:ppe} and \eqref{eq:hpe}, the observation
error dynamics are given by
\begin{equation}
\dot{\tilde{w}}(t)  = \opA_m \tilde{w}(t) +\tilde{\alpha}(t)f(w(t)),~~
\tilde{y}(t) = \opC \tilde{w}(t),~~\tilde{w}(0) = 0
\label{eq:obserr}
\end{equation}
We recall that $\opC$ is bounded.

We start by proving a bound on $\tilde{w}(t)$ that relies only on the boundedness of 
$\tilde{\alpha}(t)$. Understandably, this is a weak bound and we will subsequently make it
stronger in the following subsections under additional assumptions. A key point is that
it does not rely on the coercivity of $\opP$ in the projection-based adaptive laws.
\begin{lem}\label{eq:twbdd}
Suppose that $\|w\|_{\mathW ,\tau} < \rho_w$ for some constant $\rho_w > 0$. Then, the 
adaptive laws in \eqref{eq:pproj} ensure that $\|\tilde{w}\|_{\mathW,\tau}$
and $|\tilde{y}(t)|$ are bounded for $t < \tau$.
\end{lem}
{\flushleft{\em Proof}}: From \eqref{eq:obserr}, note that
$$
\|\tilde{w}\|_{\mathW,\tau} \leq \|\staropT(\tau)\|\|\tilde{\alpha}f(w)\|_{\mathW,\tau}
$$
Notice that $\tilde{\alpha}(t)$ is bounded for all $t$ due to the projection-based laws.
Furthermore, from Lemma~\ref{lem:lipschitz}, it follows that $\|f(w)\|_{\mathW,\tau}$ is bounded.
Since $\mathcal{T}$ is exponentially stable and since all other terms on the RHS are bounded, it follows
that $\|\tilde{w}\|_{\mathW,\tau}$ is bounded. Since the output operator $\opC$ is bounded, it follows that $|\tilde{y}(t)|$ is bounded for $t < \tau$.
$\blacksquare$

In the subsequent sections, we will strengthen the bounds on
$\tilde{w}$ and $\tilde{y}$. In particular, coercivity of 
$\opP$ will play an essential part in strengthening the
bounds on $\tilde{w}$. We will show that it is possible to 
obtain stronger bounds on $\tilde{y}$ (but not necessarily
$\tilde{w}$) in the absence of coercivity.

\subsection{Case 1: $\opA_m$ permits a 
coercive $\opP$}
We start with the case where $\opA_m$ permits a coercive solution to the Lyapunov equation. This result is a combination of those 
in \cite{nat12} and \cite{par18dac}.

\begin{lem}\label{lem:obserr}
Suppose that $\|w\|_{\mathW ,\tau} < \rho_w$ for some constant $\rho_w > 0$. Suppose that a coercive solution $\opP$ exists for 
\eqref{eq:lyap} and is used in the projection operator \eqref{eq:pproj}. Then, we have that all of the following 
terms are uniformly bounded for $t < \tau$: (i) the total observation
errors $\|\tilde{w}(t)\|_{\ltwo }$ and $|\tilde{y}(t)|$; (ii) the observation errors $\|\tilde{w}_p(t)\|_{\ltwo }$ and $|\tilde{y}_p(t)|$ 
for the particular half, and (iii) the observation errors $\|\tilde{w}_h(t)\|_{\ltwo }$ and $|\tilde{y}_h(t)|$ for the homogeneous half.
Moreover, the bounds can be made arbitrarily small by increasing $\gamma$.
\end{lem}
{\flushleft{\em Proof}}: We start by proving the bounds for the total observer error. We consider the 
Lyapunov function
\begin{equation}
V(t) = \langle \tilde{w} (t),\,\opP\tilde{w} (t) \rangle
+ \frac{1}{\gamma} \tilde{\alpha}(t)^{\top}\tilde{\alpha}(t) \label{eq:lyapunov}
\end{equation}
where the choice of $\opP > 0$ is explained in Sec.~\ref{sec:obsdyn}.

Differentiating the Lyapunov function gives
\begin{eqnarray}
\hspace{-12mm}\nonumber && \dot{V}(t) = \langle \opP\tilde{w} ,\,\opA _m 
\tilde{w}  \rangle + \langle \opA _m \tilde{w} ,\, \opP\tilde{w} \rangle \\
\hspace{-12mm} & & + 2\langle \opP\tilde{w} ,\, \tilde{\alpha}(t)f(w) \rangle
\!+\! \frac{1}{\gamma}\tilde{\alpha}(t)^{\top}\dot{\hat{\alpha}}(t)
\label{eq:lyap4}
\end{eqnarray}
Using~\eqref{eq:pproj} and the properties of the projection operator in Lemma~\ref{lem:lavr}, it follows that
\begin{equation}
\dot{V} \leq -\langle \tilde{w} ,\,\mathQ \tilde{w} \rangle
\label{eq:dotV}
\end{equation}

Since $\mathQ $ is boundedly invertible,
exists a constant $\lambda_p > 0$ satisfying
$$
\dot{V} \leq -\lambda_p \langle \tilde{w},\,\opP\tilde{w}\rangle
$$
Substituting into \eqref{eq:dotV}, and by adding and subtracting $\tilde{\alpha}^\top\tilde{\alpha}$ with suitable scaling, we get
$$
\dot{V} \leq -\lambda_p V + \frac{\lambda_p}{\gamma} \tilde{\alpha}(t)^{\top}\tilde{\alpha}(t)
$$
Since $\|w \|_{\ltwo } < \rho$, and $\tilde{\alpha}$ is bounded,
it follows that there exists constant $\rho_1 > 0$, which is
independent of $\gamma$, such that
\begin{equation}
V(t) \leq V(0)e^{-\lambda_p t} + \frac{\rho_1}{\lambda_p \gamma}
(1 - e^{-\lambda_p t})
\label{eq:Vtexp}
\end{equation}
Since $\tilde{w}(0) = 0$, we have that 
\begin{equation}
V(0) = \frac{1}{\gamma}\tilde{\alpha}(0)^\top\tilde{\alpha}(0)
\label{eq:V0exp}
\end{equation}
Using the coercivity of $\opP$, we deduce that 
$$
\|\tilde{w} (t)\|_{\ltwo } \leq \frac{\mu_{p,1} + \mu_{p,2}e^{-\lambda t}}{\sqrt{\gamma}} \leq \frac{\mu_{p,1} + \mu_{p,2}}{\sqrt{\gamma}},
$$
where the constant $\mu_{p,1}$ and $\mu_{p,2}$ depend on 
$\opP$ and $\rho_1$. A
similar bound for $\tilde{y}$ follows from the fact that $\opC$ 
is bounded. Clearly, the bounds can be made arbitrarily small by increasing $\gamma$. 

The proof for the boundedness of $\|\tilde{w}_p(t)\|_{\ltwo}$ 
and $|\tilde{y}_p(t)|$ for $t < \tau$ is identical to that for $\|\tilde{w}\|_{\ltwo}$ and $|\tilde{y}(t)|$. 
This is because the error equation for $\tilde{w}_p$ is identical to \eqref{eq:obserr}, except 
with $\tilde{w}$ therein replaced by $\tilde{w}_p$. Thereafter, we infer the bounds on 
$\|\tilde{w}_h(t)\|_{\ltwo}$ and $|\tilde{y}_h(t)|$ for $t < \tau$ using the triangle inequality.
This completes the proof.
$\blacksquare$

\subsection{Case 2: $\opA_m$ does not permit a coercive $\mathcal{P}$}
In this section, we consider the case where a 
coercive solution to \eqref{eq:lyap} cannot be
found. We prove two results here; informally speaking, 
these are either weaker results for the same set of assumptions as 
earlier, or equally strong results under stronger assumptions on the
system.

The first result is motivated by \cite{cur03} (Theorem 2 therein). We use the 
conditions of the Kalman-Yakubovich-Popov (KYP) lemma to derive a strong
bound on $|\tilde{y}(t)|$, similar to Lemma~\ref{lem:obserr}. The KYP lemma is used
routinely when dealing with output feedback problems, as in \cite{cur03}. In our paper,
it provides a way to deal with non-coercive settings when its conditions are met.

\begin{thm}\label{thm:cdi}
Consider the observer error dynamics \eqref{eq:obserr} and let $\|w\|_{\mathW,\tau} < \rho_w$ be suitably bounded.
Suppose that there exists a constant $\mathcal{F} \in \mathcal{L}(\dom(\opA_m),\ltwo)$, $\opP\in \mathcal{L}(\ltwo)$ with $\opP \geq 0$, $\mathQ :\dom(\mathcal{A}_m)\to \ltwo$ with $\langle z,\,\mathQ z\rangle \geq \epsilon_Q \|z\|_{\ltwo}^2$ for all $z\in \dom(\mathcal{A}_m)$ and an operator $\mathcal{E}\in\mathcal{L}(\mathbb{Z},\mathbb{R})$ such that 
\begin{eqnarray} \label{eq:kyp}
&& \mathcal{A}_m^{\ast} \opP z + \opP \mathcal{A}_m  z = -\mathcal{F}^\ast \mathcal{F}z -\mathQ z\\
\nonumber && \mathcal{E}\opP z = \opC z
\end{eqnarray}
for all $z\in \mathcal{D}(\mathcal{A}_m)$.
Then, $\|P^{1/2}\tilde{w}\|_{\ltwo}$ and $|\tilde{y}|$ are bounded and, moreover, the
bound can be made arbitrarily small by increasing $\gamma$.
\end{thm}
{\flushleft{\em Proof:}} The proof is a continuation of that for Lemma~\ref{lem:obserr}.
Since $\langle \mathcal{F}z,\,\mathcal{F}z \rangle \geq 0$, we recover \eqref{eq:Vtexp} and \eqref{eq:V0exp} to obtain
$$
V(t) \leq \frac{1}{\gamma}\left(\left(\tilde{\alpha}(0)^\top\tilde{\alpha}(0) - \frac{\rho_1}{\lambda_p}\right)e^{-\lambda_p t} + \frac{\rho_1}{\lambda_p}\right)
$$
which, via $V(t) \geq \langle P^{\frac{1}{2}}\tilde w(t), P^{\frac{1}{2}}\tilde w(t) \rangle$, implies that
\begin{equation}
\|P^{\frac{1}{2}}\tilde w(t)\|_\ltwo \leq \sqrt{\frac{1}{\gamma}} \sqrt{\left(\left(\tilde{\alpha}(0)^\top\tilde{\alpha}(0) - \frac{\rho_1}{\lambda_p}\right)e^{-\lambda_p t} + \frac{\rho_1}{\lambda_p}\right)}
\label{eq:P12w}
\end{equation}
Notice that the term on the RHS is bounded, and can be made arbitrarily small by reducing 
$\gamma$.

Since $\tilde{y} = \opC \tilde{w}$, we get using~\eqref{eq:kyp}
$$
|\tilde{y}(t) | = |\opC \tilde{w}(t)|=\|\mathcal{E}P^{\frac{1}{2}}\opP^{\frac{1}{2}} \tilde{w}(t)\|_{\ltwo} \leq \|\mathcal{E}\opP^{\frac{1}{2}}\| \|P^{\frac{1}{2}}\tilde w(t)\|_{\ltwo}
$$
Since $ \|\mathcal{E}\opP^{\frac{1}{2}}\|$ is bounded, we conclude using \eqref{eq:P12w} that
$|\tilde y(t)|$ is bounded and the bound can be made arbitrarily small by increasing $\gamma$.
$\blacksquare$

\begin{remark}
If the operator $\mathcal{E}$ in Thm.~\ref{thm:cdi} exists, it must satisfy the condition that 
$(\opA_m,\,\mathcal{E})$ form a controllable pair. For the conditions of the KYP lemma to be satisfied,
$(\opA_m,\,\mathcal{E},\,\opC)$ must satisfy a strictly positive real (SPR) condition.
\end{remark}

Next, we show the almost asymptotic convergence of $\tilde{y}$ (the output of the observer error dynamics) to $0$
for more general cases when the stronger assumptions of Thm.~\ref{thm:cdi} cannot be met.
\begin{thm}\label{thm:noncoercive1}
Suppose that $\|w\|_{\mathW,\tau} < \rho_w$ for some 
$\tau > 0$ and some constant $\rho_w > 0$. Then, the 
adaptive laws in \eqref{eq:pproj} ensure that 
$\int_0^t |\tilde{y}(s)|^2\,ds$ is bounded for $t < \tau$.
Furthermore, if the solution to \eqref{eq:big} exists for 
all $t$ and $\|w\|_{\mathW} < \rho_w$,  then 
$\tilde{y}$ converges to $0$ almost asymptotically
in the sense of Definition~\ref{defn:almostasymp}. 
\end{thm}
{\flushleft{\em Proof}}: We start by defining a state $v \in \mathbb{R}$ whose dynamics is defined via
$$
\frac{d}{dt}\left(\frac{v^2}{2}\right) = \tilde{y}^2,~~v(0) = 0
$$
Restricting $v$ to satisfy $v \geq 0$, this equation has a well-defined solution for all $t$.

Recall that $\mathcal{I}$ is the identity operator 
(on $\ltwo$). Let $\epsilon_1$ be an arbitarily small number such that $\mathcal{I} > \epsilon_1 \opC^\ast \opC$.
We define a Lyapunov function
$$
V = V_0 + \epsilon_1 \frac{v^2}{2}
$$
where $V_0$ is the same Lyapunov function as in \eqref{eq:lyapunov}, and $\opP$ is chosen to satisfy \eqref{eq:lyap} with $\mathQ = \mathcal{I}$, the identity
operator. Differentiating with respect to time, we get
$$
\dot{V} \leq -\langle \tilde{w},\,\tilde{w}\rangle + \epsilon_1 \tilde{y}^2 < 0
$$
Hence, $V(t)$ is bounded for $t < \tau$ and it follows that $v(t)^2$ is bounded for $t<\tau$. 

If the solution to \eqref{eq:big} exists for all $t$, it follows that $\|\tilde{y}\|_{\mathcal{L}_2}$ 
is bounded. Lemma~\ref{lem:almostasymp} then implies that $\tilde{y} \to 0$ in the almost asymptotic sense
of Definition~\ref{defn:almostasymp}.
$\blacksquare$

\section{Performance and Stability}\label{sec:stab}
\subsection{Stability}
In this section, we assert the boundedness of the control input and the stability of the closed-loop system. These results, and their proofs, are identical to those in our prior work \cite{nat12,par18dac}. These results are not altered by the 
lack of a coercive solution to the Lyapunov function

We start by asserting the boundedness of $\hat{y}_p$ in \eqref{eq:ppe}.
\begin{lem}\label{lem:yhatpbound}
Suppose that $\|w \|_{\mathW ,\tau} \leq \rho_w$ for some $\rho_w > 0$. Then, there 
exist constants $\delta_0$ and $\delta_1$ such that
$\|\hat{y}_p\|_{\linf,\tau} \leq \delta_0 + \delta_1\|w \|_{\mathW ,\tau}$.
\end{lem}

The boundedness of $\hat{y}_p$ allows us to assert that the control input $\ctrlu(t)$, given by 
\eqref{eq:upt} and \eqref{eq:pbvp}, is bounded. Let $u_r = -H_C p(t)$, the second term on the RHS of
\eqref{eq:upt}. Let $H(s) = H_C(sI-H_A)^{-1}H_B$.
\begin{lem}\label{lem:u}
Let $\|w \|_{\mathW ,\tau} < \rho_w$ for some $\tau$ and $\rho_w > 0$. Then,
the control input $\ctrlu(t)$ is bounded and a $C^1$ function of time
for $t < \tau$. Moreover, 
there exist constants 
$\delta_{0w} \equiv \delta_{0w}(H(s),\rho)$,  $\delta_{0r}\equiv \delta_{0r}(H(s),\rho)$ 
and $\delta_{0u}\equiv \delta_{0u}(H(s),\rho)$ such that
$\|u_r\|_{\linf,\tau} \leq \delta_{0w}\|w \|_{\mathW,\tau} + \delta_{0r}\|r\|_{\linf,\tau} + \delta_{0u}$. 
\end{lem}

Finally, we assert the stability of the complete closed-loop system, in the sense of 
$\mathW$-boundedness of signals, using the following small gain.
\begin{assm}[Small-gain condition] \label{assm:smallgain} We assume that there exists a constant $\rho_w$, an arbitrarily small $\epsilon_s > 0$, and a stable strictly proper 
$H(s)$ such that the following inequality is satisfied:
$$
\frac{M\rho_0 + \staropTnorm(\nu_2(\rho_w) \!+\! \delta_{0r}\|r\|_{\linf} \!+\! \delta_{0u})}{1 - \staropTnorm(\nu_1(\rho_w) + \|\opB \|\delta_{0w})}\leq \rho_w - \epsilon_s
$$
where the constants have been defined in Lemmas~\ref{lem:lipschitz} and \ref{lem:u}.
\end{assm} 

\begin{thm}\label{thm:main1}
The closed-loop system \eqref{eq:iomc}, \eqref{eq:ppe}, \eqref{eq:hpe}, \eqref{eq:pproj}, \eqref{eq:upt} and \eqref{eq:pbvp} is bounded-input-bounded-state stable in the sense of $\mathW$ if Assumption~\ref{assm:smallgain} is satisfied. Moreover, the solution exists for all time
and $\|w\|_{\mathW} < \rho_w$.
\end{thm}

\subsection{Reference signal tracking}
Ideally, we would design $u(t)$ to ensure that $\hat{y}_h(t)$ in \eqref{eq:hpe} tracks $\sigma(t)$ in \eqref{eq:sigma}. Guarantees on the tracking error between $y(t)$ and $r(t)$ depend, therefore, on the provable bounds on $\tilde{y}(t)$.
Notice that Theorems~\ref{thm:cdi} and \ref{thm:noncoercive1} provide relatively strong bounds, albeit of different natures, on $|\tilde{y}|$. In particular, Theorem~\ref{thm:noncoercive1} shows that it $\hat{y}_h$ tracks $\sigma$ asymptotically, then $y$ tracks $r$ almost asymptotically.

However, in neither of these cases (Theorems \ref{thm:cdi} and \ref{thm:noncoercive1}) is it 
possible to guarantee a strong bound on $\|\tilde{w}(t)\|_{\ltwo}$. This implies that bounds on the {\em transient characteristics} of the output tracking error cannot be asserted, unlike in the case where $\opP$ is coercive. 

\subsection{Impact on Model-Following}
In $\lon$ adaptive control, unlike MRAC, there is no explicitly prescribed reference model. Instead, we define an {\em auxiliary reference system}
\begin{eqnarray}
\hspace{-7mm}\nonumber && \dot{w}_{\rm ref}(t) \!=\! \opA_m w_{\rm ref}(t) \!+\! \opB u_{\rm ref,r}(t) \!+\! f(w_{\rm ref}),~w_{\rm ref}(0) \!=\! w(0) \\
\hspace{-7mm}&& y_{\rm ref} \!=\! \opC w_{\rm ref} 
\label{eq:auxref}
\end{eqnarray}
where $f(\cdot)$ is assumed to be known. For the auxiliary
system, we don't need state observers for
the homogeneous and particular halves. Instead we write their dynamics as
\begin{eqnarray}
\nonumber \dot{w}_{\rm ref,h} &=& \opA_m w_{\rm ref,h} + \opB u_{\rm ref,r} \\
\dot{w}_{\rm ref,p} &=& \opA_m w_{\rm ref,p} + f(w_{\rm ref}) \label{eq:refhp}
\end{eqnarray}
and calculate $u_{\rm ref,r}$ as
\begin{eqnarray}
\hspace{-5mm}\nonumber && u_{\rm ref,r} = -H_c p_{\rm ref}(t),~~\dot{p}_{\rm ref}(t) = H_A p_{\rm ref}(t) + H_B \sigma_{\rm ref}(t) \\
\hspace{-5mm}&& \sigma_{\rm ref}(t) = r(t) - y_{\rm ref,p}(t)
\end{eqnarray}
Note that this equation is similar to \eqref{eq:upt}.

We define the model-tracking error $e_w = w - w_{\rm ref}$, and the error $e_p = p - p_{\rm ref}$.
It can be checked that the dynamics of these variables are given by
\begin{eqnarray}
\nonumber \dot{e}_w &=& \opA_m e_w - \opB H_c e_p + f(w) - f(w_{\rm ref})  \\
\dot{e}_p & = & H_A e_p - H_B \opC e_w - H_B \tilde{y}_p \label{eq:ewep}
\end{eqnarray}
Subject to the small gain condition in Assumption~\ref{assm:smallgain}, \eqref{eq:ewep} represents a stable system driven by $\tilde{y}_p$.
This allows us to assert the following result.
\begin{prop}\label{prop:reftrack}
Suppose that that the closed-loop system is well-posed and bounded for all time. Then, if $\opA_m$ permits a coercive solution $\mathcal{P}$ to \eqref{eq:lyap} or if the conditions of Theorem~\ref{thm:cdi} are satisfied, the error $\|e_w\|_{\mathW}$ 
is bounded and the bound can be made arbitrarily small by increasing $\gamma$.
\end{prop}

The above proposition does not cover the case of Thm~\ref{thm:noncoercive1}. 
It is not possible to provide strong guarantees on the transient model-following
error in the absence of stronger bounds on $\|e_w\|_{\mathW}$.

\begin{remark}[Implications for MRAC]
In traditional model reference adaptive control (MRAC), adaptive laws are designed by considering the error between the actual system and a reference 
model, rather than between the actual system and an observer. Equation \eqref{eq:obserr} is sufficiently representative of the
dynamics of the model-tracking error. Since the lack of a coercive solution to \eqref{eq:lyap} prevents us from deriving a strong bound on $\tilde{w}$, it is not possible to guarantee that the transient response of the actual system matches that of the reference model. This is significant for MRAC because the primary role of the reference model is to specify desirable closed-loop transient response characteristics.
\end{remark}

\section{Conclusion}
In this paper, we examined adaptive control problems where the Lyapunov equation used
to derive adaptive laws and closed-loop performance guarantees does not permit a coercive
solution. We showed how, in the absence of coerciveness, it is still possible to provide limited
guarantees on the tracking error. We showed, in particular, that is generally possible to show that
the tracking errors decay in a weakly asymptotic manner. Under extra assumptions resembling 
those in the KYP lemma, we derived tracking performance guarantees closer to
the case where the Lyapunov equation permits a coercive solution. We demonstrated the effect
of the lack of coercivity on the nature of the provably guarantees for the transient response
of the closed-loop system.

\bibliography{Bibliography}
\bibliographystyle{plain}

\end{document}